\documentclass[twocolumn,aps,prc,showpacs,preprintnumbers,amsmath,amssymb,10pt]{revtex4-1}

\usepackage{graphicx}
\usepackage{dcolumn}
\usepackage{bm}
\usepackage{multirow}

\begin{document}

\title{Does the $\alpha$ cluster structure in light nuclei persist through the fusion process?}

\author{J. Vadas}
\author{T.~K. Steinbach}
\author{J. Schmidt}
\author{Varinderjit Singh}
\author{C. Haycraft}
\author{S. Hudan}
\author{R.~T. deSouza}
\email{desouza@indiana.edu}
\affiliation{%
Department of Chemistry and Center for Exploration of Energy and Matter, Indiana University\\
2401 Milo B. Sampson Lane, Bloomington, Indiana 47408, USA}%

\author{L.~T. Baby}
\author{S.~A. Kuvin}
\author{I. Wiedenh\"{o}ver}
\affiliation{
Department of Physics, Florida State University, Tallahassee, Florida 32306, USA}%

\date{\today}

\begin{abstract}
\begin{description}
\item[Background] Despite the importance of light-ion fusion in nucleosynthesis, a limited amount of 
data exists regarding the de-excitation following fusion for such systems.
\item[Purpose] To explore the characteristics of $\alpha$ emission associated with the decay of light fused systems at low
excitation energy.  
\item[Method] Alpha particles were detected in coincidence with evaporation residues (ER) formed by the fusion of 
$^{18}$O and $^{12}$C nuclei. Both $\alpha$ particles and ERs were identified on the basis of their energy and time-of-flight. 
ERs were characterized by their energy spectra and angular distributions while the $\alpha$ particles were characterized by their
energy spectra, angular distributions, and cross-sections.
\item[Results] While the energy spectra and angular distributions for the $\alpha$ particles are well reproduced by statistical model
codes, the measured cross-section is substantially underpredicted by the models. Comparison with similar systems reveals that the 
fundamental quantity for the $\alpha$ cross-section is E$_{\mathrm{c.m.}}$ and not the excitation energy of the fused system. 
\item[Conclusion]The enhancement in the measured $\alpha$ cross-section as compared to the statistical model codes and its dependence 
with E$_{\mathrm{c.m.}}$ suggest that a coupling between pre-existing $\alpha$ cluster structure and the collision dynamics is reponsible 
for the observed $\alpha$ cross-section. 
\end{description}
\end{abstract}

\maketitle
Nuclear fusion is a phenomenon of considerable significance from both the fundamental and societal perspective. Synthesis of the elements in both 
stellar \cite{Penionzhkevich10,Carnelli14} and non-stellar \cite{Zagrabaev12,Back14,Yanez14} environments is 
principally governed by nuclear fusion. Attempts to synthesize superheavy 
elements at the limits of stability rely on fusion reactions \cite{Zagrabaev12}. 
Not only do fusion reactions provide the path by which both existing 
and potentially new elements are synthesized, but they also provide access to
an enormous release of energy. In addition to powering stellar cores,  
it has recently been proposed that nuclear fusion reactions
in the outer crust of an accreting 
neutron star fuel the tremendous 
energy release observed in X-ray 
superbursters \cite{Strohmayer06,Horowitz08,Horowitz09a,Horowitz09b}. With an energy release of 10$^{42}$ ergs, an X-ray 
superburst releases in just a few hours the energy output of our sun over approximately a decade. Beyond their occurrence in nature, 
fusion reactions are also of practical importance. Fusion 
weapons represent the largest terrestial energy release achieved 
by human beings to date. 
Moreover, the quest to harness the
sustained energy release of fusion remains the focus of considerable effort 
\cite{NRC,Moses10}.
Due to the important role fusion reactions play, they have been intensively 
studied both 
experimentally and theoretically for several decades.

For many systems, fusion involves 
the amalgamation of two nuclei into a compound nucleus which no longer
retains a memory of the identity or structure of the colliding
nuclei. As the two nuclei fuse, 
both binding 
energy and incident kinetic energy are converted into intrinsic excitation and
spin. At energies near the Coulomb barrier, the resulting compound nucleus, 
characterized by its spin and excitation energy, de-excites by emitting 
neutrons, protons, $\alpha$ particles, and $\gamma$ rays.
To describe this de-excitation of the compound
nucleus a statistical framework is typically invoked \cite{Weisskopf37,Hauser52}. 
The defining features of the de-excitation process 
are the energy spectra and angular distributions
of the emitted particles along with their cross-sections. 
Although this perspective of fusion reactions, namely the complete equilibration
of the projectile and target nuclei followed by their statistical decay
has largely been successful,
exceptions have been noted \cite{Nagashima86}.
In these cases, it has been noted that entrance channel effects are observable.
To test this survival of entrance channel effects in fusion reactions, we
investigate the collision of light nuclei with well
established $\alpha$ cluster structure \cite{Johnson08,vonOertzen10}.
The extent to which this pre-existing cluster structure survives the
fusion process can be probed  
by examining $\alpha$ particle emission as a function of incident energy.   
In this paper we examine $\alpha$ emission in the reaction 
$^{18}$O + $^{12}$C for E$_{\mathrm{c.m.}}$ = 6.5 to 14 MeV.

The experiment was conducted at Florida State University where a beam of $^{18}$O
ions was accelerated to energies between E$_{\mathrm{lab}}$ = 16.25~MeV  and 36~MeV using the FN tandem and 
pulsed at a frequency of 12.125~MHz.
After optimizing the beam optics, the beam intensity was decreased 
to 1.5-4$\times$10$^5$~p/s
to faciltate comparison with future experiments using 
low intensity radioactive 
beams \cite{Steinbach14}. 

In the experimental setup the beam first passed through
an upstream ExB microchannel plate 
(MCP) detector designated US MCP. In this detector, passage of the 
$^{18}$O ions through a secondary emission carbon foil produced
a fast timing signal \cite{Steinbach14}. 
The beam subsequently impinged on  
a second MCP detector designated TGT MCP approximately 1.3 m 
downstream of the US MCP. 
The 93 $\mu$g/cm$^2$ 
thick carbon foil in the TGT MCP 
served both as the target for the experiment as well as 
a secondary emission foil 
for this MCP. 
Measurement of the time-of-flight (TOF) between the two MCPs allowed rejection of beam particles 
scattered or degraded prior to the target as well as provided a direct measure of the
number of beam particles incidenit on the target. The fast timing signal of the TGT MCP 
was also used to measure the TOF for reaction products.

In the angular range 
4.3$^\circ$ $\le$ $\theta_{\mathrm{lab}}$ $\le$ 11.2$^\circ$ 
reaction products were detected 
using a 
segmented, annular silicon detector which provided both an energy and 
fast timing signal \cite{deSouza11}. The detector used in this experiment was
a new design fabricated by Micron Semiconductor designated S5. The detector which
was nominally 220 $\mu$m thick consisted of sixteen
pie-shaped sectors on its ohmic surface. On its junction side, the detector was 
segmented into six concentric rings subdivided into four quadrants. 
The segmentation of this design was optimized for the kinematics associated with
the study of low energy fusion reactions.
Reaction products were also detected in the angular range 
12$^\circ$ $\le$ $\theta_{\mathrm{lab}}$ $\le$ 23$^\circ$ using another annular silicon 
detector (design S1), 300 $\mu$m thick, situated closer to the target. 
This detector was similar to the S5
detector previously described but had sixteen concentric rings spanning the angular range
which are sub-divided into quadrants.
Due to 
the kinematics of the reaction, the angular range subtended by 
these detectors resulted in a high geometric efficiency, 65$\%$-80$\%$, for detection of fusion residues.
Further details on the operating performance of these detectors and the experimental setup are
described in Refs. \cite{Steinbach14}, \cite{SteinbachPRC14}.

\begin{figure}[]
\includegraphics[width=8.6cm]{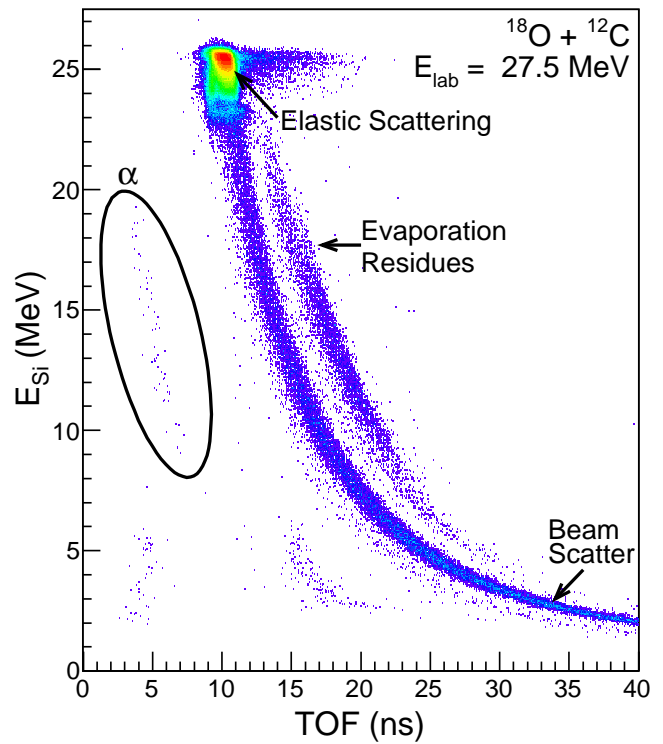}
\caption{(Color online) Energy versus time-of-flight spectrum of reaction products 
with 4.3$^\circ$ $\le$ $\theta_{lab}$ $\le$ 11.55$^\circ$. Color represents yield on a logarithmic 
scale.}
\label{fig:etof}
\end{figure}

A typical energy vs. time-of-flight (ETOF) spectrum measured is presented 
in Fig.~\ref{fig:etof} where the energy corresponds to the energy 
deposited in the silicon detector while the time-of-flight is the time difference 
between the target MCP and the silicon detector. The prominent feature in the spectrum is the 
peak associated with elastically scattered particles located at E$_{\mathrm{Si}}$ = 25 MeV and a TOF of
approximately 10 ns. Extending from this peak to lower energies is a locus of points
that exhibit a characteristic energy-TOF relationship. This locus corresponds to scattered beam
particles and has a total intensity of approximately 2$\%$ of the elastic intensity.
Situated at longer TOF than the beam scatter is a clear island corresponding to the detection of
nuclei with A$>$18. 
Located at shorter times than the beam scatter are two 
distinct islands. Located between E$_{\mathrm{Si}}$ $\approx$ 8 MeV and 20 MeV and a TOF of 4-7 ns 
is a locus corresponding to the detection of $\alpha$ particles. The expected correlation between energy
and TOF is qualitatively manifested for these particles. 
Calculation of the ETOF associated with A=4 confirms this assignment.
At lower deposited energy,
E$_{\mathrm{Si}}$ $<$ 6 MeV and a TOF of 4-5 ns an island corresponding to protons is also observed.

The region of 
Fig.~\ref{fig:etof} associated with A$>$18 is due to  
fusion of $^{18}$O nuclei with $^{12}$C nuclei. 
The resulting $^{30}$Si nuclei de-excite via emission of neutrons, protons, $\alpha$ 
particles producing evaporation residues. Detection of the evaporation residues provides a direct
measure of the fusion cross-section.

\begin{figure}[]
\includegraphics[width=8.6cm]{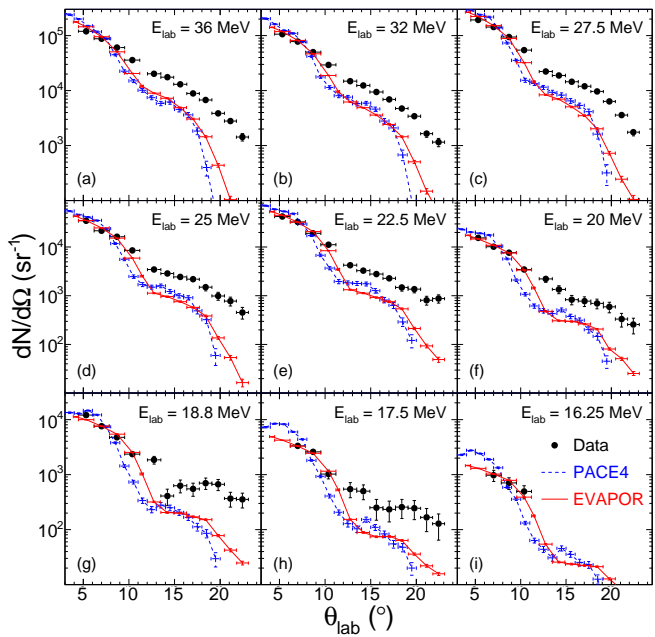}
\caption{(Color online) Angular distribution of evaporation residues in the laboratory frame for 
different bombarding energies
for $^{18}$O + $^{12}$C. Solid symbols depict the experimental angular distribution while 
the solid and dashed curves indicate the angular distributions predicted by the statistical
model codes EVAPOR and PACE4 respectively. The model angular distributions have been normalized to the
experimental data over the angular range 4.3$^\circ$ $\le$ $\theta_{\mathrm{lab}}$ $\le$ 11.2$^\circ$. 
}
\label{fig:angdist}
\end{figure}

Presented in Fig.~\ref{fig:angdist} 
is the laboratory angular distribution of 
evaporation residues for incident energies E$_{\mathrm{lab}}$=16.25 MeV to 36 MeV. 
Also shown are the evaporation residue angular distributions predicted by the statistical model 
codes EVAPOR \cite{evapor} (solid red line) and PACE4 \cite{Gavron80} (dashed blue line), which employ a Hauser-Feshbach formalism 
to describe the de-excitation of the fusion product. 
At all energies the
yield for evaporation residues decreases with increasing laboratory angle. 
Closer examination of
the angular distributions reveals that the distributions have a two component nature 
that can be qualitatively understood in the following context.
De-excitation of the fusion product via single or few nucleon emission will impart less
transverse momentum to the recoiling evaporation residue resulting in an angular
distribution that is peaked at smaller angles. In contrast, emission of an 
$\alpha$ particle will result
in a larger transverse momentum for the evaporation residue and as a result an angular
distribution that is peaked at larger angles. 
The small angle component of these distributions are well described by the statistical model codes, but the large 
angle component is significantly underpredicted.

\begin{figure}[]
\includegraphics[width=8.6cm]{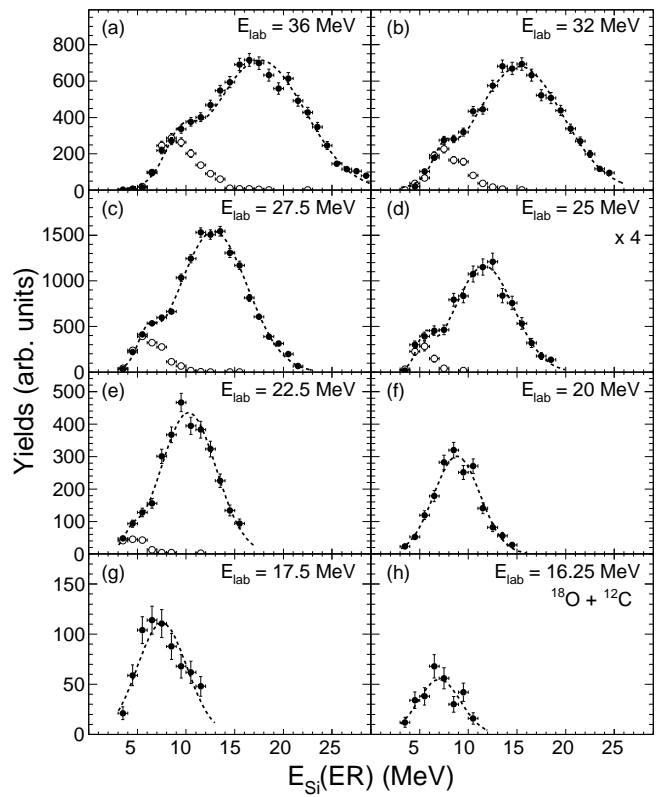}
\caption{Solid symbols depict the distribution of deposited energies in the Si detector for evaporation residues 
at different bombarding energies.
Open symbols correspond to the same quantity for which evaporation residues are coincident with $\alpha$ particles.
Open symbols have been scaled by a factor of two for clarity.
The dashed line corresponds to a two gaussian fit.}
\label{fig:eres}
\end{figure}

The energy distributions of evaporation residues are shown in Fig.~\ref{fig:eres} for different
incident energies. It should be noted that the distributions presented 
correspond to the energy deposited in the silicon detector.
As the atomic number of the residues is not known the energy measured in the
silicon detector has not been corrected for the energy loss in the target or the entrance dead
layer of the silicon detector. If one assumes, consistent with statistical model 
calculations, that the evaporation residues are predominantly Si and Al nuclei, then this 
energy loss correction
is typically of the order of 1 to 1.5 MeV. At the five higher energies a 
clear indication of a bimodal distribution 
is observed. Qualitative examination of the shape of these energy 
distributions indicates that 
the total distribution is dominated by the yield of the high energy component. This observed 
distribution can be well described by the sum
of two gaussians as shown by the two gaussian fit indicated
by the dashed 
line. 
For E$_{\mathrm{lab}}$ $\le$ 20 MeV only a single component distribution is observed corresponding to the
higher energy component present at higher beam energies.

One possible origin of the two component nature of the energy distributions visible in 
Fig.~\ref{fig:eres} is different de-excitation pathways for the excited $^{30}$Si nucleus, namely
$\alpha$ emission as compared to nucleon emission. This conclusion is also consistent 
with the angular distributions observed in 
Fig.~\ref{fig:angdist}. 
To investigate if this hypothesis is correct,
we constructed the energy distribution of evaporation residues selected on the coincident
detection of an $\alpha$ particle in the angular range
4.3$^\circ$ $\le$ $\theta_{\mathrm{lab}}$ $\le$ 23$^\circ$. The results are presented as the open 
symbols in 
Fig.~\ref{fig:eres}. All the residue energy distributions 
coincident with an $\alpha$ particle 
are single peaked with 
maxima at E$_{\mathrm{Si}}$ = 6-9 MeV.
The fact that the $\alpha$ gated residue energy 
distributions are peaked
at essentially the same location as the mean value of the low energy component 
and have comparable widths, provides strong evidence that 
the low energy component in Fig.~\ref{fig:eres} is associated with $\alpha$ emission.
The reduction of the average energy of
the evaporation residue is understandable since the $\alpha$ particle is detected at 
forward angles hence the recoil
imparted to the evaporation residue lowers its energy.

\begin{figure}[]
\includegraphics[width=8.6cm]{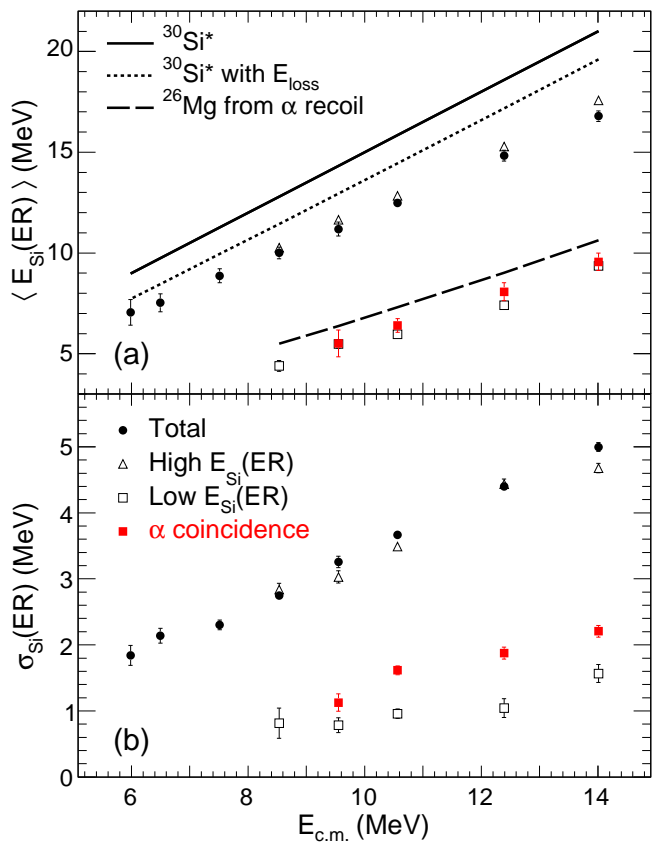}
\caption{(Color online) Top panel: Average energy deposited in the Si detector by fusion 
residues as a function of the 
available energy in the center-of-mass (solid circle). The mean energy 
extracted 
for the low and high energy components from the fits shown in Fig.~\ref{fig:eres} 
are represented by
the open squares and triangles respectively, while the red closed squares 
correspond to residues in coincidence with $\alpha$ 
particles. 
The solid line represents the energy of the excited compound nucleus 
for complete fusion. The dotted line represents the compound nucleus energy 
after energy loss in both the target and Si dead layer. The dashed line represents
the average energy deposited by a $^{26}$Mg nucleus following emission of an 
$\alpha$ particle.
Bottom panel: Widths, $\sigma$, associated with the mean values shown in the top panel.}
\label{fig:ewidth_ecm}
\end{figure}

A quantitative perspective of the trends associated with the low and high 
energy component is
examined in Fig.~\ref{fig:ewidth_ecm}. In the upper panel of the figure one 
observes that 
for both the high energy (open triangles) and low energy (open squares) 
components the
average laboratory energy of the residue, $\langle$E$_{\mathrm{Si}}(ER)$$\rangle$, increases 
essentially linearly with the
incident energy E$_{\mathrm{c.m.}}$. 
As expected, the trend for the total distribution (filled circles) 
follows that of the high energy
component since the yield of the high energy component dominates the yield of the 
total distribution. 
The trend of the $\alpha$ gated residue energy distributions (solid red squares) unsurprisingly 
follows that of the low energy component, quantitatively demonstrating that the low energy residues 
are associated with $\alpha$ emission. 
At the lowest incident energies measured, the low energy of these
evaporation residues emphasizes the need for low detection thresholds. 
The linear trend observed for the average energies of the residues can be understood 
as the change of the kinematics of the reaction with increasing incident energy.
To quantitatively assess this dependence we have calculated the average laboratory energy of the $^{30}$Si fusion 
product as a function of E$_{\mathrm{c.m.}}$ and indicate the result as the solid line in
Fig.~\ref{fig:ewidth_ecm}.
To investigate the discrepancy between the measured values for the evaporation residues (solid circles) and that calculated 
for the $^{30}$Si (solid line) we have calculated the energy a 
$^{30}$Si nucleus would possess after it passes through
the target and front dead layer of the Si detector. 
The impact of the target and front dead layer of the Si detector on the 
detected energy of the $^{30}$Si has been calculated using the energy loss program 
SRIM \cite{SRIM} and the result is depicted
as the dotted line.
Also shown in Fig.~\ref{fig:ewidth_ecm} is the $\langle$E$_{\mathrm{Si}}(ER)$$\rangle$ 
associated with a $^{26}$Mg nucleus resulting from the 
$\alpha$ decay of $^{30}$Si. The $\alpha$ emission is assumed to be isotropic with 
both the $\alpha$ particle and evaporation residue 
detected in the experimental setup. The overall agreement of the dashed line with 
the low energy component bolsters the conclusion that
the low energy component is associated with emission of an $\alpha$ particle.

In the lower panel of 
Fig.~\ref{fig:ewidth_ecm} the trends associated with the widths of the 
high and low energy components of the total distributions as well as the $\alpha$ gated distributions
are shown. The widths of both components of the total distributions increase linearly with
E$_{\mathrm{c.m.}}$ from 1.8 MeV to 5 MeV in the former case and from 0.8 to 1.6 MeV in 
the latter case.  
While the mean values of the $\alpha$ gated distributions are in good agreement 
with those of the low energy component, the widths of the $\alpha$ gated distributions are
systematically slightly larger.

\begin{figure}[]
\includegraphics[width=8.6cm]{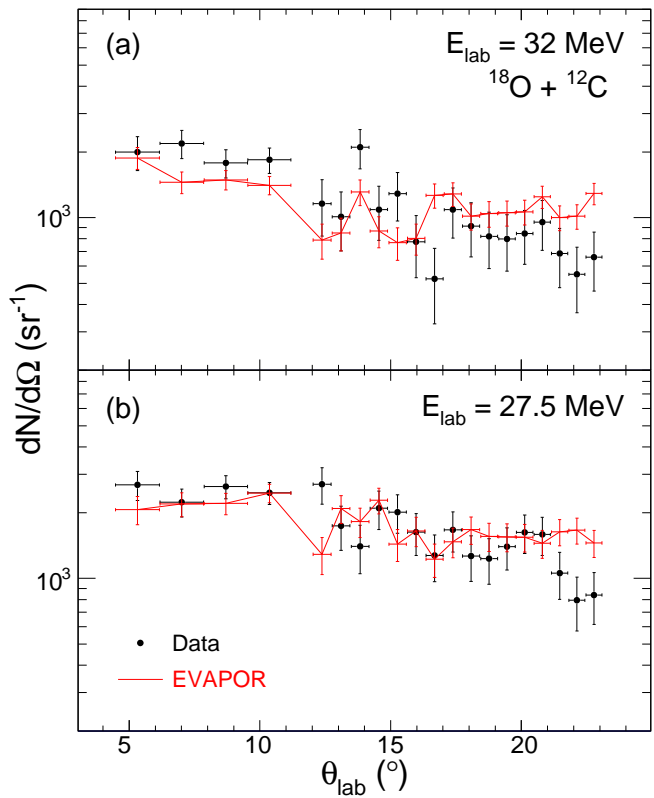}
\caption{(Color online) Angular distributions of $\alpha$ particles in the laboratory frame
at E$_{\mathrm{lab}}$=32 MeV and 27.5 MeV. The predictions of the EVAPOR model are 
indicated as a
solid (red) line.}
\label{fig:athetaphi}
\end{figure}

We next examine the measured angular distributions of $\alpha$ particles 
to ascertain if they exhibit the characteristics of statistical emission 
from a compound nucleus. The $\alpha$ particles are identified based
upon their position in the energy-TOF spectrum. Shown in 
Fig.~\ref{fig:athetaphi} are the $\alpha$ particle angular distributions 
at two incident energies
along with the predictions of the EVAPOR statistical model code normalized to the data.
The general trend observed is that the differential yield of 
$\alpha$ particles, dN/d$\Omega$,
decreases slightly with increasing angle. This forward peaking can be 
understood as being due to the 
center-of-mass momentum 
of the compound nucleus. The measured angular distributions are in relatively good agreement 
with the EVAPOR predictions as evident in the figure.

\begin{figure}[]
\includegraphics[width=8.6cm]{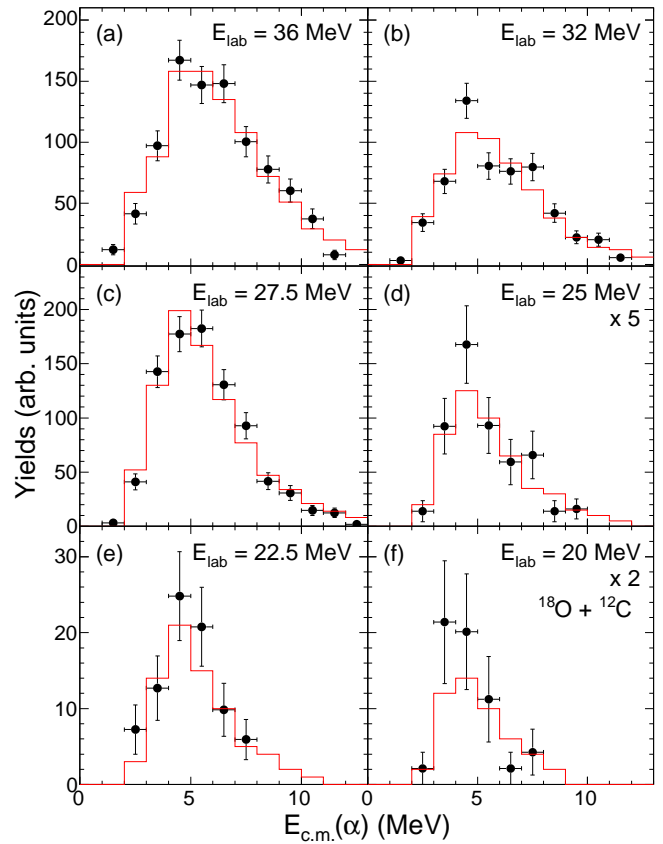}
\caption{(Color online) Energy of $\alpha$ particles in the center-of-mass frame for different bombarding energies.
The solid (red) line depicts the prediction of the statistical model code EVAPOR. The predictions have been normalized
to the experimental ones in the energy range shown.}
\label{fig:alphaecm}
\end{figure}

Having established that the $\alpha$ angular distribution 
is consistent with statistical decay from the compound nucleus and
plays a non-negligible role in the de-excitation of the fusion
product, we directly examine the energy spectra of these emitted particles. 
Shown in Fig.~\ref{fig:alphaecm} are the energy distributions of $\alpha$ particles 
detected in the angular
range 4.3$^\circ$ $\le$ $\theta_{\mathrm{lab}}$ $\le$ 23$^\circ$. 
To facilitate comparison with a statistical model, the energy of the 
$\alpha$ particle has been
transformed into the center-of-mass frame of the system and the resulting 
distributions are shown 
in Fig.~\ref{fig:alphaecm} 
along with the EVAPOR predictions. 
As is evident in the figure, the statistical model provides a reasonably good 
description of the measured energy distributions of emitted $\alpha$ particles.

\begin{figure}[]
\includegraphics[width=8.6cm]{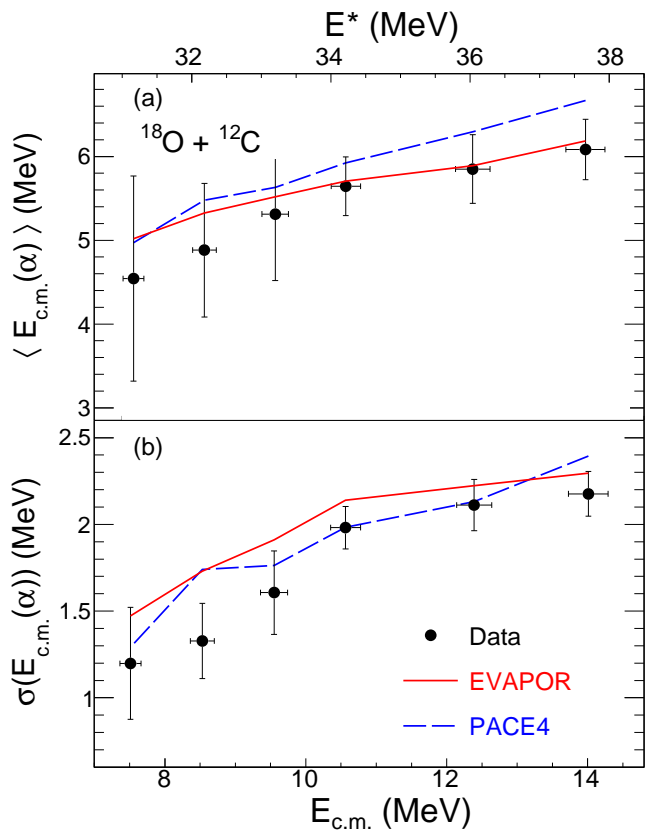}
\caption{(Color online) Top panel: Average energy of $\alpha$ particles in the center-of-mass frame as a function of the
 available energy in the center-of-mass (solid circle). 
The solid (red) line represents the average energy 
predicted by the statistical model code, EVAPOR. The dashed (blue) line represents the average energy predicted by PACE4.
Bottom panel: Widths, $\sigma$(E$_{\mathrm{c.m.}}$($\alpha$)), associated with the mean values shown in the top panel.
}
\label{fig:alphaecm_trends}
\end{figure}

In order to make a more quantitative analysis of the measured distributions 
and provide more detailed comparison with statistical models, we extract
the first and second moments of the distributions presented in Fig.~\ref{fig:alphaecm} and
examine the dependence of these quantities on E$_{\mathrm{c.m.}}$ in 
Fig.~\ref{fig:alphaecm_trends}. 
In the upper panel of Fig.~\ref{fig:alphaecm_trends} one observes that
$\langle$E$_{\mathrm{c.m.}}$($\alpha$)$\rangle$ 
increases with increasing incident 
energy, E$_{\mathrm{c.m.}}$, both for the experimental data and the model predictions. 
For reference, the excitation energy, E$^{*}$, of the compound nucleus is displayed on the scale 
above the top panel. 
The error bars for the experimental data are defined by the statistics of the measurement.
The results of the EVAPOR and PACE4 calculations are presented as the solid and dashed lines respectively. 
The overall increasing trend of the first moment, $\langle$E$_{\mathrm{c.m.}}$($\alpha$)$\rangle$,
observed in the experimental data is reasonably reproduced by both models. EVAPOR is in better agreement
with the experimental data than PACE4, which slightly overpredicts 
$\langle$E$_{\mathrm{c.m.}}$($\alpha$)$\rangle$ at all energies by approximately 0.5 MeV.  
This deviation between PACE4 and the experimental data increases with increasing E$_{\mathrm{c.m.}}$. 
While for the lower energies the statistical model predictions lie within the statistical uncertainties 
of the experimental measurement, for the two highest incident energies the statistical uncertainty 
is less than the deviation between the PACE4 model predictions and the measured values.  
Presented in the lower panel of Fig.~\ref{fig:alphaecm_trends} is the dependence 
of the second moment of the energy distributions,
$\sigma$(E$_{\mathrm{c.m.}}$($\alpha$)) on E$_{\mathrm{c.m.}}$. The experimental widths increase from 1.2 MeV
at the lowest energies to 2.2 MeV at the highest E$_{\mathrm{c.m.}}$. 
In the case of the second moment, good agreement between the
PACE4 predictions and the measured widths is observed. In contrast
to the PACE4 predictions, EVAPOR predicts slightly lower values for the first moment which are in better
agreement with the experimental measurement. However, in the case of the second moment EVAPOR slightly overpredicts the
experimentally measured values.

In a statistical framework, two factors contribute to the 
$\langle$E$_{\mathrm{c.m.}}$($\alpha$)$\rangle$ namely the temperature of the emitting nucleus 
and the Coulomb barrier associated with the $\alpha$ emission. 
As the second moment is primarily sensitive to 
the temperature of the emitting system, the larger disagreement of the PACE4 statistical model with the first moment suggests
that the Coulomb barrier associated with $\alpha$ emission might be slightly lower than that calculated by the 
statistical model.
A sensitive probe of the Coulomb barrier 
is the emission probability of a charged particle. We therefore examine the 
$\alpha$ particle emission cross-section as a function of E$_{\mathrm{c.m.}}$ and 
compare the results to the predictions of the statistical models. 

\begin{figure}[]
\includegraphics[width=8.6cm]{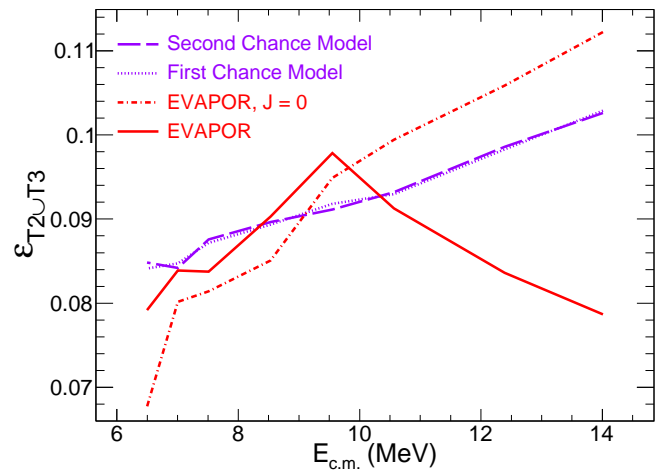}
\caption{(Color online) Efficiency for detection of an $\alpha$ particle in coincidence with an evaporation residue
 in the experimental setup as determined by the EVAPOR model. 
Also shown is the prediction of a zero spin kinematic model described in the text.}
\label{fig:alphaeffic}
\end{figure}

In order to extract the $\alpha$ emission cross-section from the measured yields, it is necessary to correct for 
the efficiency of the experimental setup. 
To determine the geometric acceptance of the experimental setup the statistical model code EVAPOR was utilized. 
In the simplest case of isotropic single $\alpha$ particle emission, two factors dominate the geometric efficiency, 
namely the center-of-mass velocity of the compound nucleus and the energy distribution of the emitted $\alpha$ particle. 
Emission of additional particles, however, imparts momentum to the evaporation residue 
which will affect the efficiency. The efficiency determined using the EVAPOR model is shown in  
Fig.~\ref{fig:alphaeffic} as a solid (red) line.
The efficiency for
detection of an $\alpha$ particle in coincidence with an evaporation residue ranges increases from 7.9$\%$ at E$_{\mathrm{c.m.}}$ = 6.5 MeV
to a maximum of 9.8$\%$ at 
E$_{\mathrm{c.m.}}$ = 9.5 MeV. A further increase in the incident energy results in a decrease of the efficiency to $\approx$7.8$\%$
at E$_{\mathrm{c.m.}}$ = 14 MeV. 
The initial increase can be understood as due to the effect of kinematic focusing.

To assess the principal factors impacting the efficiency we constructed a simple model. This model accounted for sequential two-body decays of
the compound system, emitting an $\alpha$ particle followed by a neutron (first chance) or a neutron followed by an $\alpha$ particle (second chance). In this model, the compound nucleus, $^{30}$Si, travelling with a velocity, {\it v}$_{\mathrm{CN}}$, along the beam direction emits 
the first particle. Isotropic emission is assumed consistent with zero spin. Momentum is conserved between the emitted particle and the resulting evaporation residue. The second particle is then emitted isotropically from the evaporation residue, and momentum is again conserved.  The products are then subjected to a software replica of the experimental setup to determine the efficiency. The resulting 
efficiency is depicted as a dotted line (first chance) and a dashed line (second chance) in Fig.~\ref{fig:alphaeffic}. At the lowest incident energies measured the simple model is in good agreement
with the efficiency calculated using EVAPOR. For incident energies E$_{\mathrm{c.m.}}$ $>$ 9.5 MeV, the simple model and EVAPOR diverge.
The divergence of the simple model and EVAPOR may signal the increasing importance of angular momentum which is absent in the simple model. At 
E$_{\mathrm{c.m.}}$= 14 MeV the maximum angular momentum is calculated to be $\approx$10$\hbar$. 
To ascertain if the angular momentum of the compound nucleus was responsible for decrease in efficiency we calculated
the efficency for compound nuclei with zero angular momentum (J=0) within the EVAPOR model. As can be seen in 
Fig.~\ref{fig:alphaeffic} for this case the efficiency increases monotonically with increasing incident energy. 
As the EVAPOR model includes the competition between 
different channels as well as the treatment of angular momentum, we utilized the efficiency determined using EVAPOR to extract the 
$\alpha$ emission cross-section.

\begin{figure}[]
\includegraphics[width=8.6cm]{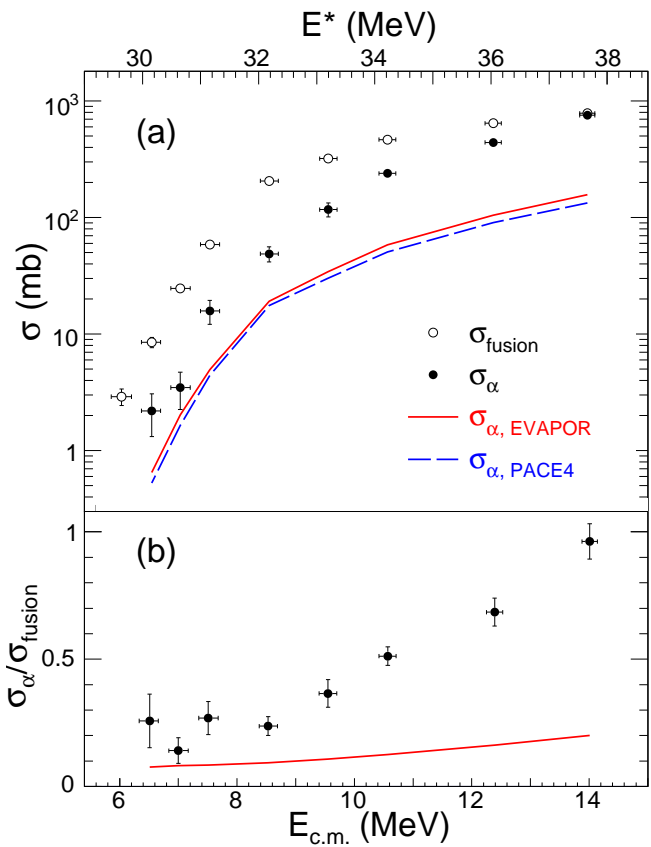}
\caption{(Color online) Top panel: Comparison of the measured $\alpha$ emission cross-sections (closed symbols) 
with the statistical model codes EVAPOR (solid red line) and PACE4 (dashed blue line). 
The total fusion cross-sections are shown as the open symbols. 
Bottom panel: Alpha emission cross-sections relative to the total fusion cross-sections as a function of 
E$_{\mathrm{c.m.}}$ for the experimental data (closed symbols) and EVAPOR (solid red line).}
\label{fig:alphaxsect}
\end{figure}

Presented in Fig.~\ref{fig:alphaxsect} is the cross-section for $\alpha$ decay following fusion of the
$^{18}$O and $^{12}$C nuclei. 
In the top panel of Fig.~\ref{fig:alphaxsect} one observes that the cross-section for $\alpha$ decay increases with 
increasing excitation energy with a shape consistent with a barrier emission process. 
Over the interval measured the $\alpha$ cross-section increases 
from approximately 2 mb to 700 mb. 
The total fusion cross-section is also shown for reference. As might be 
qualitatively expected, at small excitation energy, E$^*$, only a relatively small fraction of the total fusion cross-section is 
associated with $\alpha$ decay. This fraction increases with increasing excitation energy. 
Also shown for comparison are the predictions of the statistical model 
codes EVAPOR (solid line) and PACE4 (dashed line). 
The cross-section predicted by the models has been obtained by utilizing the relative probability for all $\alpha$ 
channels and the experimentally measured total fusion cross-section.
While the models exhibit the same qualitative behavior as observed experimentally, 
both EVAPOR and PACE4 substantially underpredict the experimentally measured cross-sections.

The dramatic increase in the relative cross-section for
 $\alpha$ emission with excitation energy and the underprediction of the statistical model codes is emphasized in the
lower panel of Fig.~\ref{fig:alphaxsect}. At the lowest excitation energies $\alpha$ emission comprises approximately 
20$\%$ of the fusion cross-section. This fraction increases rapidly becoming essentially unity by an E$^*$ of 38 MeV. Over the 
excitation energy interval measured, EVAPOR only predicts an increase in the relative $\alpha$ emission from $\approx$10$\%$ to 
20$\%$. From the upper panel of Fig.~\ref{fig:alphaxsect}
it is clear that the result for PACE4 would be essentially the same.
The discrepancy between the experimental data and the statistical model predictions is twofold. Not only 
do the statistical model calculations underpredict the magnitude of the relative $\alpha$ particle emission, but they
underpredict the rate at which $\alpha$ particle emission increases with E$_{\mathrm{c.m.}}$. This result suggests that factors
other than those considered in the statistical model calculations play a significant role in the $\alpha$ particle emission.

While the dramatic increase in the $\alpha$ emission cross-section with incident energy and the underprediction of the statistical 
model codes is remarkable, it should be noted that a hint of this result was 
already evident in the angular distribution of evaporation residues
presented in Fig.~\ref{fig:angdist}. As observation of residues at large laboratory angles is directly related to the
emission of an $\alpha$ particle, the failure of the statistical model codes to reproduce the yield of evaporation residues
at large angles suggests the underprediction of $\alpha$ emission. Although the energies of the emitted $\alpha$ particles are reasonably
reproduced by the statistical model codes and in particular EVAPOR, the models underpredict the measured $\alpha$ cross-section. 
Moreover, the magnitude of the underprediction increases with increasing incident energy. At the highest incident energy measured the
statistical model code EVAPOR underpredicts the measured $\alpha$ cross-section by a factor of approximately five. 

\begin{figure}[]
\includegraphics[width=8.6cm]{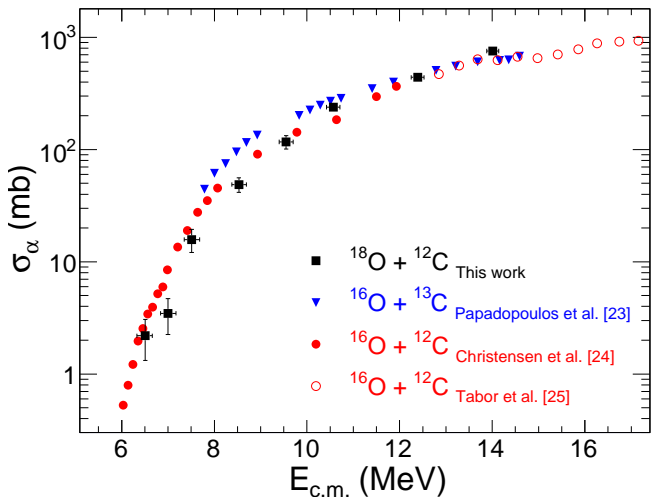}
\caption{(Color online) Dependence of the $\alpha$ emission cross-section on E$_{\mathrm{c.m.}}$ for several O + C systems.}
\label{fig:alphaxslit}
\end{figure}

Presented in Fig.~\ref{fig:alphaxslit} are the measured $\alpha$ excitation functions for 
$^{16}$O + $^{12,13}$C \cite{Papadopoulos86,Christensen77,Tabor77} along with the present data. As the incident 
energy increases from E$_{\mathrm{c.m.}}$ = 6 MeV 
to E$_{\mathrm{c.m.}}$ = 17 MeV, $\sigma_\alpha$ increases from $<$1 mb to $\approx$1000 mb. The similarity of the
$\alpha$ cross-section for the different systems presented as a function of E$_{\mathrm{c.m.}}$ is striking.
Though slight differences in the shape of the excition functions in the range 
8 $\le$ E$_{\mathrm{c.m.}}$ $\le$ 11 MeV exist, the excitation functions largely follow a common trend.
If the $\alpha$ cross-section primarily depended on the 
excitation energy, E$^*$, of the fused system, then a common trend would be observed for the
different systems as a function of E$^*$. However, the Q-value for the three systems
is significantly different ranging from 16.76 MeV for $^{16}$O + $^{12}$C 
to 23.65 MeV for $^{18}$O + $^{12}$C. Consequently, when the dependence of the
$\alpha$ cross-section on E$^*$ is examined, the different systems are displaced relative to each other
by approximately the difference in their Q-values. This result implies that 
E$_{\mathrm{c.m.}}$ and not E$^*$ is the fundamental quantity driving $\alpha$ emission in these systems.

One can speculate as to why E$_{\mathrm{c.m.}}$ is the relevant quantity for $\alpha$ emission in these fused 
systems. It is well established that nuclei such as $^{12}$C and $^{16}$O have an $\alpha$ cluster
structure. Even for the neutron-rich nucleus $^{18}$O significant experimental evidence for
an $\alpha$ cluster structure exists \cite{Johnson08,vonOertzen10}. If this $\alpha$ cluster structure in the
entrance channel is not eliminated in the fusion process, it could manifest itself as enhanced $\alpha$
emission relative to the statistical model. Evident in the lower panel of Fig.~\ref{fig:alphaxsect}
is the growth of the relative probability for $\alpha$ emission with E$_{\mathrm{c.m.}}$. This growth substantially 
exceeds that predicted by the statistical model calculations suggesting that the pre-existing $\alpha$ cluster
alone does not explain the magnitude of $\alpha$ emission for large values of E$_{\mathrm{c.m.}}$. From this we surmise
that the collision dynamics coupling to the inherent $\alpha$ cluster structure is responsible for
the large cross-section of $\alpha$ particles observed. While it may be tempting to consider the survival and amplification 
of the pre-existing $\alpha$ cluster structure as simply a ``pre-equilibrium'' component, it should be recalled
that the $\alpha$ particle angular distribution is consistent with that of the statistical model predictions. 
Hence, the lifetime of this ``pre-equilibrium'' process is long on the timescale of the rotational period of the fused system.

In summary, we have measured evaporation residues  and $\alpha$ particles produced  
in the reaction $^{18}$O + $^{12}$C at
16.25 MeV $\le$ E$_{\mathrm{lab}}$ $\le$ 36 MeV and examined their angular distributions, energy spectra,
as well as cross-sections. Evaporation residues exhibit
a two-component angular distribution.
The smaller angle component 
can be understood as 
associated with nucleon emission from the fused system and
is reasonably well 
described by a statistical model code (EVAPOR). 
In contrast, the yield of the 
larger angle component which is 
associated with the emission
of $\alpha$ particles 
is significantly underpredicted by the model
indicating that $\alpha$ emission is enhanced relative to the predictions of the
statistical model code.
While the 
angular distributions and energy spectra of the emitted $\alpha$ particles
are in good agreement with the statistical model code predictions, the measured $\alpha$ cross-section
far exceeds the predicted cross-section. This enhancement of the cross-section increases from a 
factor of two at E$_{\mathrm{c.m.}}$ = 7 MeV to  a factor of nearly five at E$_{\mathrm{c.m.}}$ = 14 MeV.
This large $\alpha$ cross-section is also observed for other light systems undergoing fusion. 
Remarkably, comparison with similar systems indicates that 
E$_{\mathrm{c.m.}}$ and not excitation energy is the quantity responsible for the $\alpha$ emission process.
This result indicates that the $\alpha$ particles are not emitted from the {\it fully} equilibrated 
compound nucleus despite the agreement of the angular distribution with the statistical model code.
Moreover, the growth of the $\alpha$ particle cross-section with increasing incident energy is revealing.
This increase of the relative
$\alpha$ cross-section with increasing incident energy can be understood as 
a coupling of the collision dynamics to pre-existing $\alpha$ cluster structure in the entrance channel.
The large increase in the observed $\alpha$ cross-section 
with increasing incident energy 
may signal the increased coupling of the entrance channel to the $\alpha$ cluster structure as the
bombarding energy increases. 
As light-ion fusion reactions and alpha cluster nuclei in particular play 
an important role in stellar nucleosynthesis, it is important to explore this 
observation further both experimentally and theoretically. On the experimental front, 
measuring  $\alpha$ emission for similar systems which lack a pronounced $\alpha$ cluster
structure in the projectile and target nuclei is necessary to determine if an $\alpha$
cluster structure in the entrance channel is necessary to observe the enhancement. 
While acquiring high quality experimental data in a systematic fashion is crucial,
a complete understanding this phenomenon will require a theoretical model capable of treating
the $\alpha$ cluster structure in the entrance channel and its coupling to the collision dynamics.

We wish to acknowledge the support of the staff at Florida State University's John D. Fox accelerator in 
providing the high quality beam that made this experiment possible. This work was supported by the 
U.S. Department of Energy under Grant No. DE-FG02-88ER-40404 (Indiana University) 
and the National Science Foundation
under Grant No. PHY-1064819 (Florida State University).

\bibliography{fusion_18O_lcp}

\end{document}